\documentclass[twocolumn,aps,11pt,tightenlines,groupedaddress,superscriptaddress]{revtex4-1}
\usepackage[english]{babel}
\usepackage{graphicx}
\usepackage{prettyref}
\usepackage{amsmath}
\usepackage{color}
\usepackage{appendix}
\newrefformat{fig}{Fig.~\ref{#1}}
\newrefformat{eq}{Eq.~(\ref{#1})}
\newrefformat{tab}{table~\ref{#1}}
\newrefformat{sec}{Sec.~\ref{#1}} \newrefformat{app}{Appendix~\ref{#1}} \newrefformat{chap}{Chapter~\ref{#1}}
\newcommand{\be}{\begin{equation}} \newcommand{\ee}{\end{equation}}
\begin{document}
\title{Prediction in complex systems: the case of the international trade network}
\author{Alexandre Vidmer}
\email{alexandre.vidmer@unifr.ch}
\affiliation{Department of Physics, University of Fribourg, Fribourg CH-1700, Switzerland}
\author{An Zeng}
\affiliation{Department of Physics, University of Fribourg, Fribourg CH-1700, Switzerland}
\affiliation{School of Systems Science, Beijing Normal University, Beijing 100875, P.R. China}
\author{Mat\'u\v s Medo}
\affiliation{Department of Physics, University of Fribourg, Fribourg CH-1700, Switzerland}
\author{Yi-Cheng Zhang}
\affiliation{Department of Physics, University of Fribourg, Fribourg CH-1700, Switzerland}
\begin{abstract}
Predicting the future evolution of complex systems is one of the main challenges
in complexity science. Based on a current snapshot of a network, link prediction
algorithms aim to predict its future evolution.
We apply here link prediction algorithms to data on the international trade between countries. This data can be represented as a complex
network where links connect countries with the products that they export. Link
prediction techniques based on heat and mass diffusion processes are employed
to obtain predictions for products exported in the future.
These baseline predictions are improved using a recent metric of country
fitness and product similarity. The overall best results are achieved with a newly
developed metric of product similarity which takes advantage of causality in the network evolution.
\end{abstract}
\maketitle

\section{Introduction}

In the past few years, the international trade network has attracted the attention of researchers from various fields, and
especially from the complex networks scientists. 
The international trade was studied for the first time under the network framework in \cite{snyder1979structural}, using a blockmodel
consisting in partitioning countries together according to their trade flows, military interventions, diplomatic exchanges, and conjoint treaty
memberships. The complex network approach was used in \cite{serrano2003topology} and showed that the international trade exhibits common features
with the World Wide Web network. It has been shown that many features of the international trade can be explained using models
based on the gravity equation \cite{tinbergen1962analysis,fagiolo2010international,bhattacharya2008international}. Recently, not only
the countries, but also their exports have been analyzed under the complex network framework.
The Product Space attempts to explain how the nations develop by projecting their exports on a 2D map, and observing how they
diffuse in the Product Space \cite{hidalgo2007product}. The Economic Complexity aims to rank products by the technological
requirements needed for a country to be able to manufacture a product, and to rank the countries by their competitiveness
\cite{hausmann2007you,cristelli2014overview}.

The prediction of quantity and price of exports in the international trade has been studied using various models \cite{hummels2005variety}
and additional information such as geographical distance between countries and common language \cite{rauch1999networks}.
By contrast, we employ here a recommendation approach that is usually applied to e-commerce systems \cite{liben2007link,lu2011link}
in order to predict what an individual country will export in the future.
The prediction of which products a country will add to its 
export basket can help to understand how countries grow. The countries' future exports are particularly 
complex to predict, as their evolution depends on many external factors, such as geographical position, diplomatic relations
between countries, available natural resources, and technologies. Nevertheless, previous studies \cite{hidalgo2007product,tacchella2012new}
showed that it is possible to measure the competitiveness of a country, estimate its future growth and even predict its future exports 
using solely the international trade data. The last aspect, personalized prediction of future exports for each country, further suffers from 
the lack of a traditional approach with conventional metrics and renowned prediction algorithms.

Recommender systems were created to filter the relevant information in information systems \cite{resnick1997recommender,adomavicius2005toward}.
For instance, an algorithm can help a person to choose which movie to watch by creating a list of most relevant movies that this particular
person might enjoy, whereas it is a tremendous task to find a movie of interest among the thousands of existing ones. Recommendation
using temporal and causal effects has recently attracted attention \cite{zhu2015consistence,koren2010collaborative}. In this paper, we use a network
representation for countries and their exports. A network is made of nodes connected by links. Nodes represent 
countries and products, and a link connect a country to a product if the country exports the product. We call this type of network \textit{bipartite},
as the nodes are formed by two disjoint sets: countries and products. This allows us to treat the problem within
a \textit{recommender system} framework \cite{lu2011link}. The approach that we use slightly differs 
from a link prediction approach because we aim to predict the future exported products for each country instead of
predicting the most likely future exports for the network as a whole.

We use a recent recommender algorithm based on diffusion \cite{zhou2010solving} as a tool for prediction.
However, as demonstrated in \prettyref{fig:1}, the international trade data fundamentally differs from online systems data on which
recommendation is typically done by the absence of preferential attachment \cite{barabasi1999emergence}. This makes it impossible to predict the future popularity
of a product from its current popularity and the predictions thus have correspondingly lower accuracy. To improve the recommendation 
accuracy, we adopt a temporal approach to devise a new definition of mutual distance between products, and couple it with the diffusion
method in order to enhance the prediction of the links. This approach is closely related to the proximity of products
\cite{hidalgo2007product}, defined as a distance between each couple of products depending on the number of co-occurrences in
countries' export baskets. Based on this distance, it was shown that the closer a product is to one of the products the country
is currently exporting, the more likely it is to be added to the country's export
basket in the future. The recommendation of movies to users was improved by the use of co-occurrences in \cite{zhu2015consistence}.
In~\cite{koren2010collaborative}, the authors studied the temporal dependence of movies ratings and used it to improve the prediction of
future ratings.

In order to improve the prediction performance of the diffusion method, we also use Economic Complexity as defined in \cite{tacchella2012new}.
The first definition of Economic Complexity was made by the use of two self-consistent linear equations \cite{hidalgo2009building,hausmann2014atlas},
which were then successfully applied to predict the long-term growth of countries' export basket. A more recent definition was given as a set of two
self-consistent non-linear equations \cite{tacchella2012new} that reflect the non-linear relations between complexity of products and
competitiveness of countries. This non-linear definition of complexity was shown to capture more information than the linear one on the countries'
hidden growth potential in a toy model of countries' exports \cite{cristelli2013measuring}.
The use of the complexity of products in the prediction process allows us to improve the diversity of the prediction, while maintaining
the same level of accuracy. In contrast with the mutual distances introduced previously, the Economic Complexity theories aim to assign each
country and product an individual score on an absolute scale.

\section{Methods}
\subsection{Data}
We use the NBER-UN dataset which was described and cleaned in \cite{feenstra2005world}.
We cleaned it further by removing aggregate categories and keeping only the countries
for which complete mutual export data are available. Products having zero total export volume for a given year while
having nonzero total export volume for the previous and the following years were removed from the dataset.
Products and countries with no entries after year 1993 were removed as well.
After the cleaning procedure, the network consists of 65 countries and 770 products. 
To decide if we consider country $i$ to be an exporter of product $\alpha$ or not, we use the Revealed Comparative
Advantage (RCA) \cite{balassa1965trade} which is defined as

\be 
\mathrm{RCA}_{i\alpha}=\frac{e_{i\alpha}}{\sum_\beta e_{j\beta}} \Bigg/ \frac{\sum_j
e_{j\alpha}}{\sum_{j\beta}e_{j\beta}},
\ee
where $e_{i\alpha}$ is the volume of product $\alpha$ that country $i$ exports in thousands of US dollars.
RCA characterizes the relative importance of a given export volume of a product
by a country in comparison with this product's exports by all other countries.
We use a bipartite network representation with two different kinds of nodes, one for countries
and one for products. All country-product pairs with values above a RCA threshold---which is set to 1---are
consequently joined by links between the corresponding nodes in
the bipartite network. Before the RCA threshold is applied, there
are 35,881 links between the countries and products. In the year 1998, 10,148 links
are above the RCA threshold, which means that 20\% of all possible links between countries and
products are present. In this paper, Latin symbols $i,j$
are used for countries and Greek symbols $\alpha,\beta$ for products. The set of links
present at year $t$ is labeled $\mathcal{C}^t$, and the set of new links that have been added between time $t-1$ and
$t$ is labeled $\mathcal{N}^t := \mathcal{C}^t \setminus \mathcal{C}^{t-1}$ (the set of new links could also be built
with a longer time step, i.e $\mathcal{N}^t := \mathcal{C}^t \setminus \mathcal{C}^{t-\Delta t}$, see \prettyref{app:1}).

Note that we omit time indices in the following sections 
for the sake of clarity and we implicitly discuss the status of the network at time $t$.
The degree values $k_i$ and $k_\alpha$, are the number of links originating at a country node $i$ 
and a product node $\alpha$, respectively.

\subsection{International trade metrics}

In many online bipartite networks, such as users and movies, users do not require any special skill to watch a movie.
Conversely, countries need capabilities in order to produce a good and then export it. And so it is in particular important to capture
the relations between items, as a country needs to build its path towards a product in order to add it to its export basket
\cite{hidalgo2009building}. We want to take into account this major difference in our prediction process by using quantities
especially developed for the international trade.
We first describe two metrics---\textit{proximity} and \textit{causality}---used to assess the relations
between products, which can be further extended to describe the relations between countries and products.
In \cite{hidalgo2007product} the proximity $\phi_{\alpha \beta}$ of products $\alpha$ and $\beta$
was introduced as the symmetrized probability that a country exports product $\alpha$, given that it exports product $\beta$.
\be 
\phi_{\alpha \beta}=\min(P(\alpha \in \mathcal{C}_i|\beta \in
\mathcal{C}_i ), P(\beta \in \mathcal{C}_i |\alpha \in \mathcal{C}_i)),
\ee
where $\mathcal{C}_i$ is the set of products exported by country $i$, 
$P(\alpha \in \mathcal{C}_i |\beta \in \mathcal{C}_i)$ is the probability of exporting product
$\alpha$ given that $\beta$ is being exported, and $\phi_{\alpha \beta}$ is the proximity of 
$\alpha$ and $\beta$. The proximity $\Phi_{i \alpha}$ of product $\alpha$ to country $i$ is defined
as the average proximity of products in $i$'s export basket to product $\alpha$
\be
\Phi_{i \alpha}= \frac{1}{k_i}\sum_{\beta \in \mathcal{C}_i} \phi_{\alpha \beta}.
\label{eq:prox_close}
\ee
Based on proximity, we propose a measure labeled causality which takes into account
the time order in which products are introduced in production, similar to the URL/HTTP prediction in \cite{sarukkai2000link}
and to the country-product time evolution analysis in \cite{zaccaria2014taxonomy}.
Causality is defined as the conditional probability that country $i$ starts to export product $\alpha$
at time $t$, given that it does export product $\beta$ at time $t$,
\be
\psi_{\alpha \beta}=P(\alpha \in \mathcal{N}^{t}_i | \beta \in
\mathcal{C}^t_i).
\label{eq:caus}
\ee
Its method of computation of is given in Appendix B.
Based on this, a modified closeness $\Psi_{i \alpha}$ between product $\alpha$ and country $i$ is defined as
the average causality between product $\alpha$ and $i$'s export basket
\be
\Psi_{i \alpha}= \frac{1}{k_i}\sum_{\beta \in \mathcal{C}_i} \psi_{\alpha \beta},
\label{eq:caus_close}
\ee
where $\psi_{\alpha \beta}$ is defined in \prettyref{eq:caus}.

The third metric assigns a score to individual countries and products instead of describing the relations
between products. In \cite{tacchella2012new}, the country-product network was studied using a set of self-consistent
equations to compute fitness of countries and complexity of products. Fitness of a country indicates its ability
to manufacture complex products, relatively to other countries, while complexity of a product indicates
the amount of technology required to produce it.
Country fitness and product complexity are defined as
\be\begin{array}{ll}
F_i^{n}=&\sum_{\alpha \in \mathcal{C}_i} Q_\alpha^{n-1} \\ Q_\alpha^{n} =& 1\big/\big(\sum_{i\in
\mathcal{C}_\alpha} 1/F_i^{n-1}\big)
\end{array}, 
\ee 
where $n$ is the current iteration, $\mathcal{C}_\alpha$ is the set of countries that export product $\alpha$, 
$F_i^{n}$ is the fitness of country $i$ at iteration step $n$ and $Q_\alpha^{n}$ is the
complexity of product $\alpha$ at step $n$. Fitness and complexity
are initialized as $F_i^{0}=Q_\alpha^{0}=1$ and normalized after each iteration so that their sum is $N$ and $M$, respectively. 
The convergence of the algorithm and its stopping condition were studied in \cite{pugliese2014convergence}.

\subsection{Link prediction}

Algorithms inspired by heat \cite{zhang2007heat} and mass \cite{zhou2007bipartite} diffusion were designed
to recommend items to users in bipartite networks \cite{zhou2010solving}. If the two methods are coupled 
together, the resulting hybrid diffusion method is one of the best performing link prediction method in
bipartite networks without explicit rating of items \cite{lu2012recommender}.

The relative weight of heat and mass diffusion is controlled by an adjustable parameter $\lambda$. Each product $\alpha$ is assigned an initial
resource $f_\alpha$, and each resource is propagated by $\mathbf{\tilde{f}}=\mathbf{W} \mathbf{f}$, where the elements of the hybrid diffusion
matrix $\mathbf{W}$ are
\be 
W_{\alpha \beta}=\frac{1}{k_\alpha^{1-\lambda} k_\beta^{\lambda}}\sum_{j=1}^{N}\frac{a_{j\alpha} a_{j\beta} }{k_j},
\label{eq:2nd} 
\ee
where $a_{i\alpha}$ is 1 if $RCA_{i\alpha}\geq 1$ and 0 otherwise. The matrix $\mathbf{W}$ corresponds to the propagation of resource through
the links of the networks, similar to either heat diffusion ($\lambda=0$) or mass spreading ($\lambda=1$). The matrix notation is an elegant
way of mixing the two diffusion processes; but its interpretation is not straightforward. We refer to the original paper \cite{zhou2010solving}
for a detailed discussion.
Elements of the initial resource vector for a given country are set to 1 for all the products
that meet the RCA threshold and zero otherwise. 

We attempt to improve the hybrid prediction method by including the causality closeness $\psi_{i \alpha}$
in the final diffusion score as
\be
f^\prime_{i \alpha} = f_{i \alpha} \Psi_{i \alpha}
\ee
where $f_{i \alpha}$ is the score that the hybrid method assigns to the link between country $i$ and product $\alpha$. We label this method
\textbf{causality+hybrid}.
In this way, recommendation gives preference to products that are considered similar to the current production of country $i$, in terms of causality.
Proximity $\Phi_{i\alpha}$ can be used in the same way, giving rise to the \textbf{proximity+hybrid} method.

In the same spirit, we attempt to improve our export predictions by including the complexity scores in the hybrid
diffusion process by multiplying the initial resource assigned to the product $\alpha$ with a factor
\be
w_{\alpha}=R_{\alpha,t}^{-\theta} \label{eq:comp},
\ee 
where $\theta$ is an adjustable parameter and $R_{\alpha,t}$ is the complexity rank of product $\alpha$ 
at time $t$ (ranks 1 and $M$ correspond to the product of highest and lowest complexity, respectively).
Note that fitness of countries as well as rankings obtained by the {\it Method of Reflections} \cite{hidalgo2009building,hausmann2014atlas} can also be used
to weight the propagation process, but the results are nearly identical to the complexity modification.
For $\theta>0$, the weights computed in \prettyref{eq:comp} give more initial resources to high complexity
items. By giving additional resources to high complexity products, a higher score is diffused to the products
that are also of high complexity, as high complexity products are exported by high fitness countries.
\subsection{Link prediction metrics}

In this section, we define six metrics to measure the quality of the prediction, three assess its precision
and three evaluate its diversity.
Recommender system assign a score to each country-product couple. 
To measure how good are the predictions based on the data from a given year, we focus on all links in the country-product network
that are absent in that year and appear in the next year (new exports). The ranking scores $r$ is obtained by ranking the
corresponding products in the score list of the corresponding countries, and compute the relative rank $r_{i\alpha }=p_{i\alpha}/(M-k_i)$,
where $p_{i\alpha}$ is the rank of product $\alpha$ in country $i$'s score list.
By averaging this quantity over all newly added links, we obtain the ranking score $r$.

The \textit{prediction list} for an individual country is
built with the top $L$ scoring products that do not meet the RCA threshold in year $t$.
where $L$ is set to 20 in this work (see \prettyref{app:3}).
A correct prediction occurs if a link is present in the prediction list at a given year, and in the network in the next year.
We denote by $D_i(L)$ the number of correct predictions for country $i$. By averaging $D_i(L)$ over countries and normalizing
it by the length of the prediction list $L$, we obtain precision $P(L)$ \cite{zhou2010solving}. Precision of
0.1 means that 10\% of the products in the top $L$ score lists are correctly predicted. 
Countries differ in size and the number of new items that they introduce in their export basket each year.
To take this variety into account, we use recall $R_i(L)=D_i(L)/N^{\mathrm{new}}_{i}$, where $N^{\mathrm{new}}_{i}$ is the number of
links originating from country $i$ in $\mathcal{N}^t$. 
Overall recall $R(L)$ is obtained by averaging $R_i(L)$ over all countries.
Recall complements precision; the two metrics are often considered simultaneously.

The diversity of the prediction lists is also an important aspect of the methods.
The first two metrics assess the degree and complexity for correctly predicted links.
Note that in \cite{zhou2010solving} these metrics were computed over all prediction lists' entries, not only the correctly
predicted ones. A subscript $c$ is added to point out this difference. For each product $\alpha$, we compute the
self-information $I_\alpha=\log_2(N/k_\alpha)$ \cite{tribus1961thermostatics}. By averaging it over all correctly predicted links for country $i$ we obtain 
$I_{c,i}(L)$, and its average over countries is novelty $I_c(L)$. The higher the score, the more items with low degree we predict.
We argue that a good prediction method should predict products of high complexity as they require more advanced technology
than low complexity ones. We further define the quantity $C_{c,i}(L)$ as the average complexity rank of products that are
correctly predicted for country $i$. Its average over each country gives the complexity metric $C_c(L)$.
We finally measure how the prediction lists differ from each other with the inter-list Hamming distance \cite{hamming1950error} (also named personalization)
$h_{ij}(L)=1-q_{ij}(L)/L$, where $q_{ij}$ is the number of common products in the prediction
lists of countries $i$ and $j$. The overall personalization $h(L)$ is obtained by averaging over all country
pairs; the values of zero and one correspond to all lists being identical and mutually exclusive, respectively.

\section{Results}
\begin{figure}[t!] \centering \includegraphics[width=1\columnwidth]{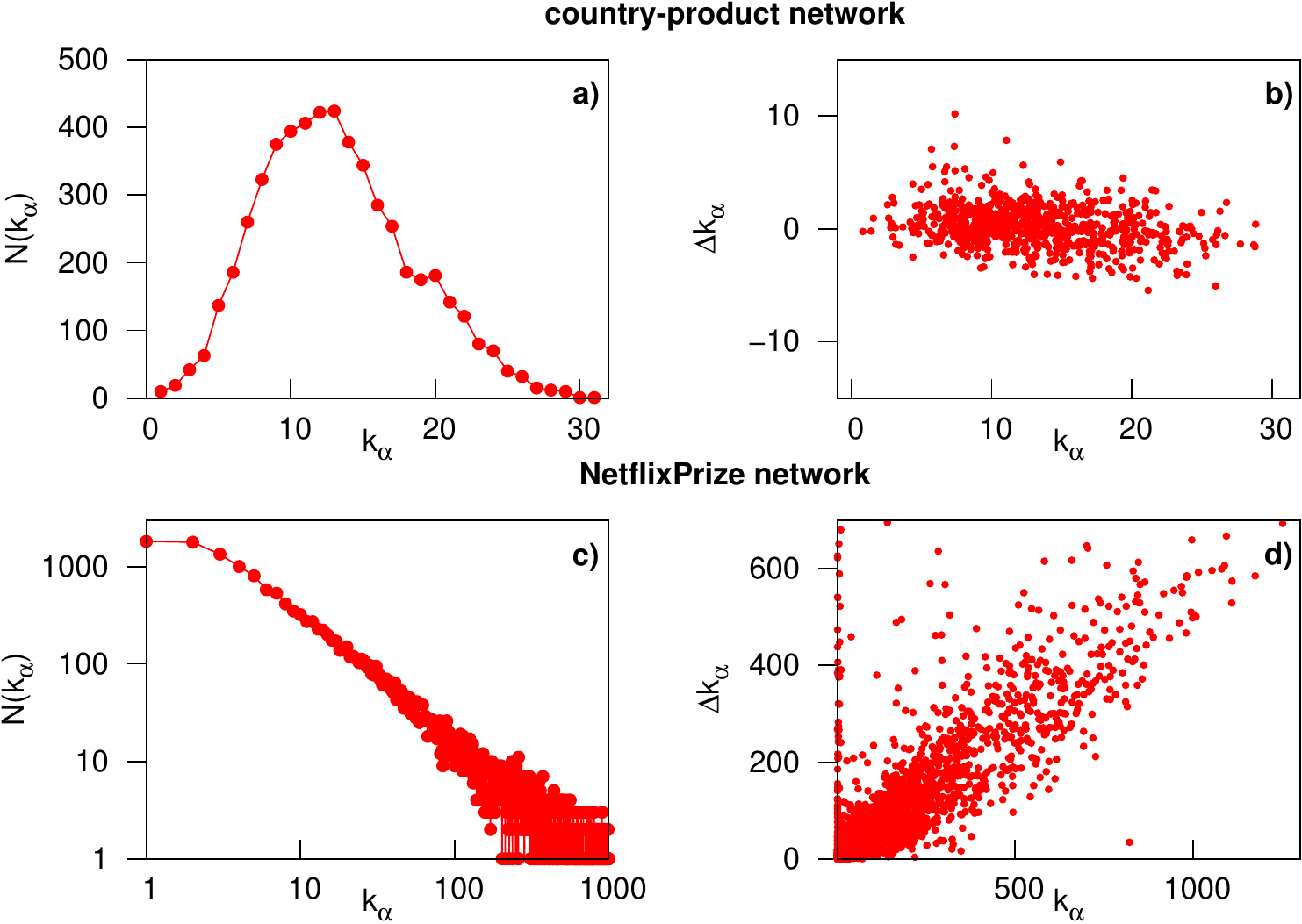} \caption{Panels
a) and c) show the item degree distribution for the country-product network and the user-movie
network, respectively. $N(k_\alpha)$ is the number of item of degree $k_\alpha$. Panels b) and d) show the relation between item 
degree increase $\Delta k_\alpha$ between two time steps and item degree $k_\alpha$. The time step is one year for the
country-product network and two hundred days for the Netflix Prize network. Random displacements by values ranging
from $-0.5$ to $0.5$ along both axes were added to reduce the extent to which the symbols overlap.}
\label{fig:1} \end{figure}

To begin our analysis, we compare the statistical properties of the country-product network with other types of real bipartite networks. 
Bipartite networks representing online systems typically consist of user- and item-nodes with connections between
them drawn when a user has collected, bought, rated, or otherwise interacted with an item. The item degree distribution
is usually broad and often exhibits a power-law shape \cite{lu2012recommender}. This is a direct consequence of 
the \textit{preferential attachment} process \cite{barabasi1999emergence},
which occurs in many real networks, such as scientific collaboration networks \cite{newman2001scientific}, metabolic networks
\cite{jeong2000large} and social networks \cite{newman2001structure}. The preferential attachment assumption is based on
the observation that high degree nodes attract additional links at a higher rate than low degree nodes.
We show in \prettyref{fig:1}a that the degree distribution on the product side of the country-product
network differs greatly from the power-law distribution shown in \prettyref{fig:1}c for the Netflix Prize network \cite{bennett2007netflix}
(Netflix Prize was a recommendation contest organized by the DVD rental company Netflix; the data consist of users and DVDs that they have rated).
As shown in \prettyref{fig:1}b and d, the degree increase per time step is also different for the two networks.
The degree increase of products in the country-product network is weakly negatively correlated with the current degree (the linear
correlation coefficient is $r=-0.28$), while for the Netflix Prize network there is a strong positive correlation
between degree increase and the current degree ($r=0.77$). As we shall soon see, this difference and the resulting high diversity
of the products that receive new links in one time step reduce the prediction accuracy in the country-product data as compared to
the accuracy in user-item data. While the preferential attachment model is a good description
of the growth in the Netflix Prize network, it is obviously not a suitable one for the country-product network.
Models based on hidden capabilities of countries and required capabilities for producing various
products seem more appropriate in this respect \cite{hidalgo2009building,cristelli2013measuring}.

\begin{figure}[t!] \centering \includegraphics[width=1\columnwidth]{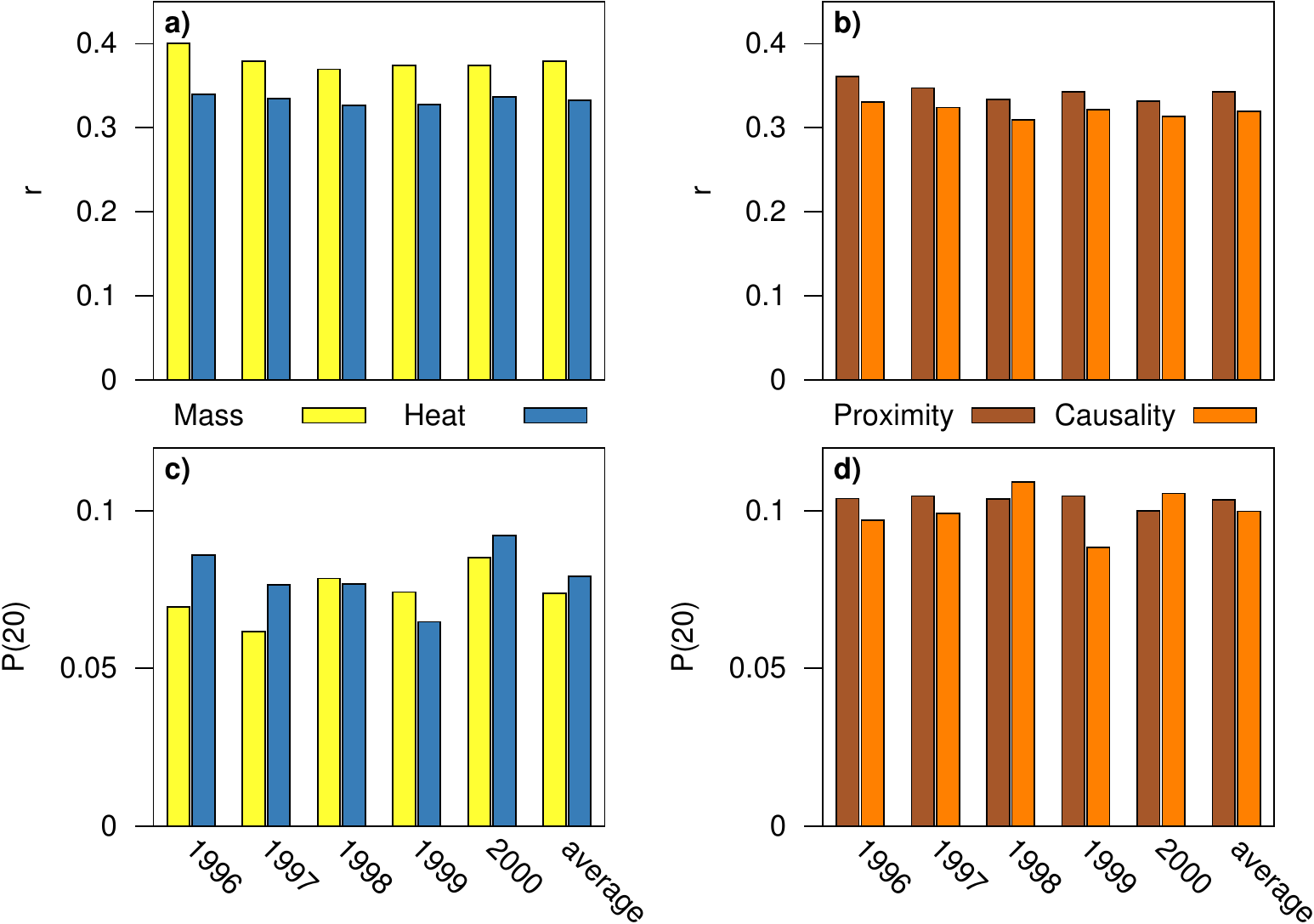} \caption{
Accuracy evaluation in terms of the ranking score (a,b) and precision (c,d)
for prediction algorithms based on diffusion (left column) and relations between the products (right column).
The year on the $x$-axis is the year for which predictions are made using the state of the network in the previous year.
}
\label{fig:2} \end{figure}

\begin{figure}[t!]
\centering
\includegraphics[width=1\columnwidth]{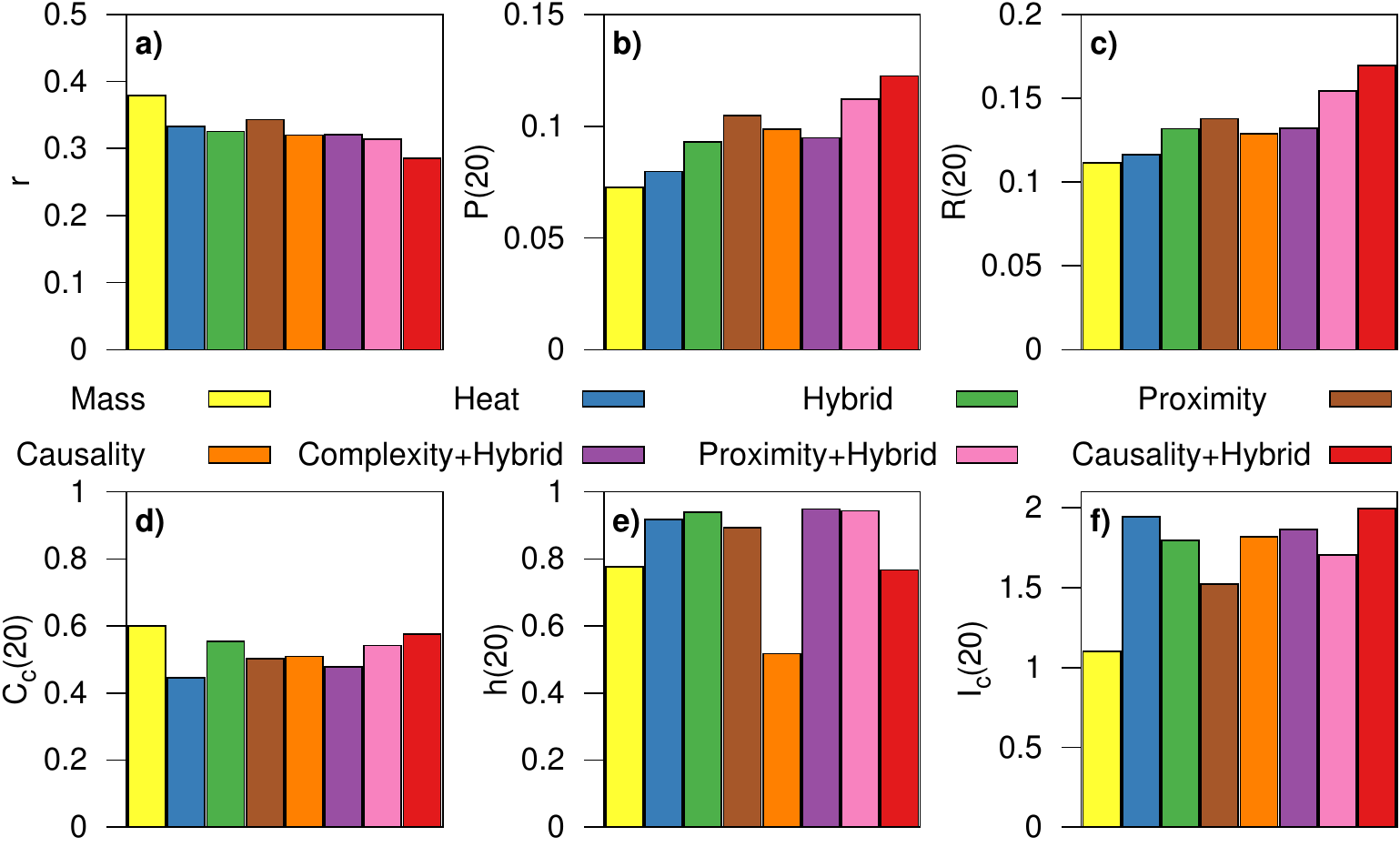}
\caption{A comparison of prediction accuracy (top row) and diversity (bottom row).
The performance results shown in this figure represent the average over results for years 1996-2000.
Methods with parameters have been optimized with respect to their ranking score.
}
\label{fig:3}
\end{figure}

\begin{figure}[t!] \centering \includegraphics[width=1\columnwidth]{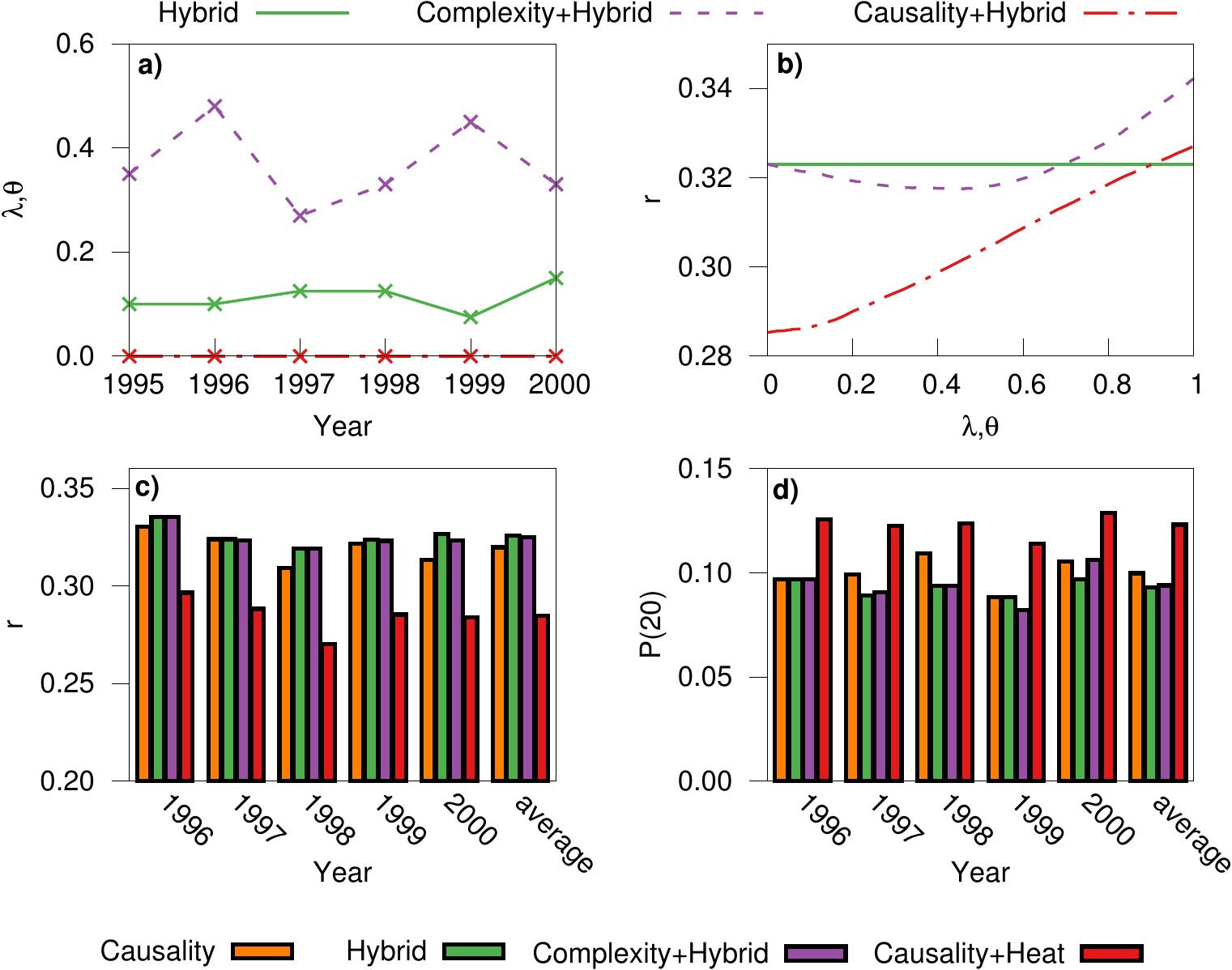} \caption{
Panel a) shows methods' optimal parameters at various years (optimization is again with respect to the ranking score).
Panel b) shows the dependence of the ranking score on the parameters for year 1998.
The results of the optimized Hybrid method are shown for comparison.
Results shown in c) and d) are made using the data in year 1995 and before to predict the new products in 1996.
Methods' parameters are here fixed based on the optimal parameters of previous year's prediction.
Performance of predictions based on causality is shown for comparison.}
\label{fig:4} \end{figure}

The performance of four basic prediction methods are compared in different years in \prettyref{fig:2}.
In user-item data, mass diffusion outperforms significantly its heat diffusion counterpart
\cite{lu2012recommender,zhou2010solving}. However, as \prettyref{fig:2}a and c show, the 
situation is very different in the country-product data: heat diffusion outperforms mass diffusion in both ranking score and precision.
The reason lies in the absence of preferential attachment in the country-product data as shown in \prettyref{fig:1}b and d. An example
using a simple model with and without preferential attachment is provided in \prettyref{app:4}.
Heat diffusion which, unlike mass diffusion, does not favor popular items \cite{zeng2012reinforcing} is thus better suited for the prediction task here.
Causality and proximity can be used alone to predict the future exports of countries (see \prettyref{fig:2}b, d).
Causality outperforms proximity in ranking score, which indicates that the temporal aspect of $\psi_{i\alpha}$ captures different
relations between products than $\phi_{i\alpha}$. At the same time, proximity outperforms causality in the precision metrics, indicating that
top-scoring products are more relevant with proximity than causality.
When optimizing the prediction performance, we found that averaging proximity
from year 1992 provides the best results, while causality benefits from a longer time period and data from as early as 1984 were used to build
the causality relations (see \prettyref{app:2}).

We compare the performance of all the methods in \prettyref{fig:3}. In agreement with
\cite{zhou2010solving}, the hybridization between mass and heat diffusion improves every accuracy metrics, as well as prediction diversity.
Compared to the hybrid method, complexity+hybrid (\prettyref{eq:comp})
slightly improves every metric, albeit at the expense of adding an additional free parameter in the prediction process.
Proximity and causality both perform similarly to the
hybrid diffusion method, without any parameters, and both have their strong points: causality yields better ranking score and
predicts products of lower degree, while proximity yields better precision and predicts products of higher complexity (\prettyref{fig:3}d).
By coupling proximity or causality with diffusion, we further improve the results. The best overall performance is obtained with 
causality+hybrid. Compared to random predictions, the precision metric is improved by a factor of 4 as compared
to a factor of 80 in the Netflix Prize network \cite{zhou2010solving}, which indicates a comparably low predictability of future links
appearing in the country-product network. Note that we can add an additional parameter in the causality+hybrid method
by exponentiating the causality score ($\psi_{i\alpha} \rightarrow \psi_{i\alpha}^\theta$, where $\theta$ is a free parameter),
resulting in an improvement of roughly 3\% on the ranking score, but a substantial improvement (around 20\%) of the personalization metric. 

The selection of parameters is an important issue for those non-parameter-free methods.
We generally use the whole dataset to optimize the parameters of our prediction methods.
To control for possible overfitting, we use an approach similar to the three-fold validation which is common in information filtering
\cite{abu2012learning,zeng2013information}. We first find the parameters for year $t$
that minimize the ranking score, and then use the optimized parameters to make the prediction for 
year $t+1$. \prettyref{fig:4}a shows that the optimal parameters are nearly constant over time for Hybrid and causality+hybrid methods.
\prettyref{fig:4}b further shows that the parameter range in which complexity and causality coupled with diffusion methods
outperforms hybrid diffusion method is quite wide. The ideal parameter of causality+hybrid method
is very close to 0, which makes the
causality+hybrid effectively a parameter-free method.
In \prettyref{fig:4}d, we set the parameters to a fixed value before the prediction.
We see that the accuracy and diversity improvements are still present for every method, and both complexity and
causality coupled with diffusion improve further those results.

\section{Discussion}
We used the hybrid diffusion algorithm \cite{zhou2010solving} one of the standard
recommendation methods in unweighted bipartite networks, to predict new links in the
country-product export network. 
Unlike for the usual user-item data, heat diffusion algorithm produced satisfactory results,
which we attributed to the growth mechanism of country-product data where preferential attachment---a 
key driving force in user-item data---is absent.
Recently developed metrics for country fitness and product complexity were used to enhance the
prediction performance. While they carry information about individual countries and products
and generally enhance prediction performance, we found relations between products---proximity and causality---to be
even more beneficial. The best overall prediction method was achieved by combining the heat diffusion
recommendation with causality scores.

In this work, we restricted our input information to the state of the country-product network obtained by
applying the RCA threshold; the detailed information on the export volume has been ignored.
If we remove this restriction and use for instance the RCA values for prediction, we can achieve high
prediction accuracy simply by ranking the products by their RCA value in the prediction list (products 
whose RCA exceeds 1 are naturally excluded because the corresponding links already exist).
This results in the ranking score $r\cong 0.15$, and precision $P\cong 0.3$, which is a significant improvement
over the best-performing causality+hybrid method which yields $r\cong 0.3$, $P\cong 0.11$. The improvement is a direct
consequence of using additional information (relative link importance quantified by the RCA metric) which is furthermore
very specific to the country-product network and thus not available in other systems to which our prediction approach may
be relevant (bipartite user-item networks where links connect items with the users who have purchased or otherwise
connected with them, for example).

The introduction of the causality score demonstrates the possibility and benefits of
using time information in the prediction process. This score can also find its use in other kinds of data,
such as user-movie data where a user who appreciated the first episode
of a series of movies is likely to watch also the second one. Apart from prediction, causality needs to be studied further to
understand what it tells us about the relations between products; for example by inspecting which products are exported before
a new product starts to be exported. The prediction in the country-product network can be further improved
external information on geography (neighboring countries may have similar capabilities) and temporal evolution of prices and export
volumes (their growth may attract new producers). Prediction approaches built on multiple models and components motivated by machine
learning \cite{bell2007lessons,koren2009matrix,jahrer2010combining} could eventually contribute to understanding the limited prediction
precision in country-product data.

\section{Acknowledgments}
This work was supported by the Swiss National Science Foundation Grant
No. 200020-143272 and by the EU FET-Open Grant No. 611272 (project
Growthcom).

\begin{appendices}
\renewcommand{\thefigure}{\thesection.\arabic{figure}}
\renewcommand{\theequation}{\thesection.\arabic{equation}}

\section{Effect of timestep on precision metrics}
\setcounter{figure}{0}
\setcounter{equation}{0}
\label{app:1}
\begin{figure}[t!]
\centering
\includegraphics[width=1\columnwidth]{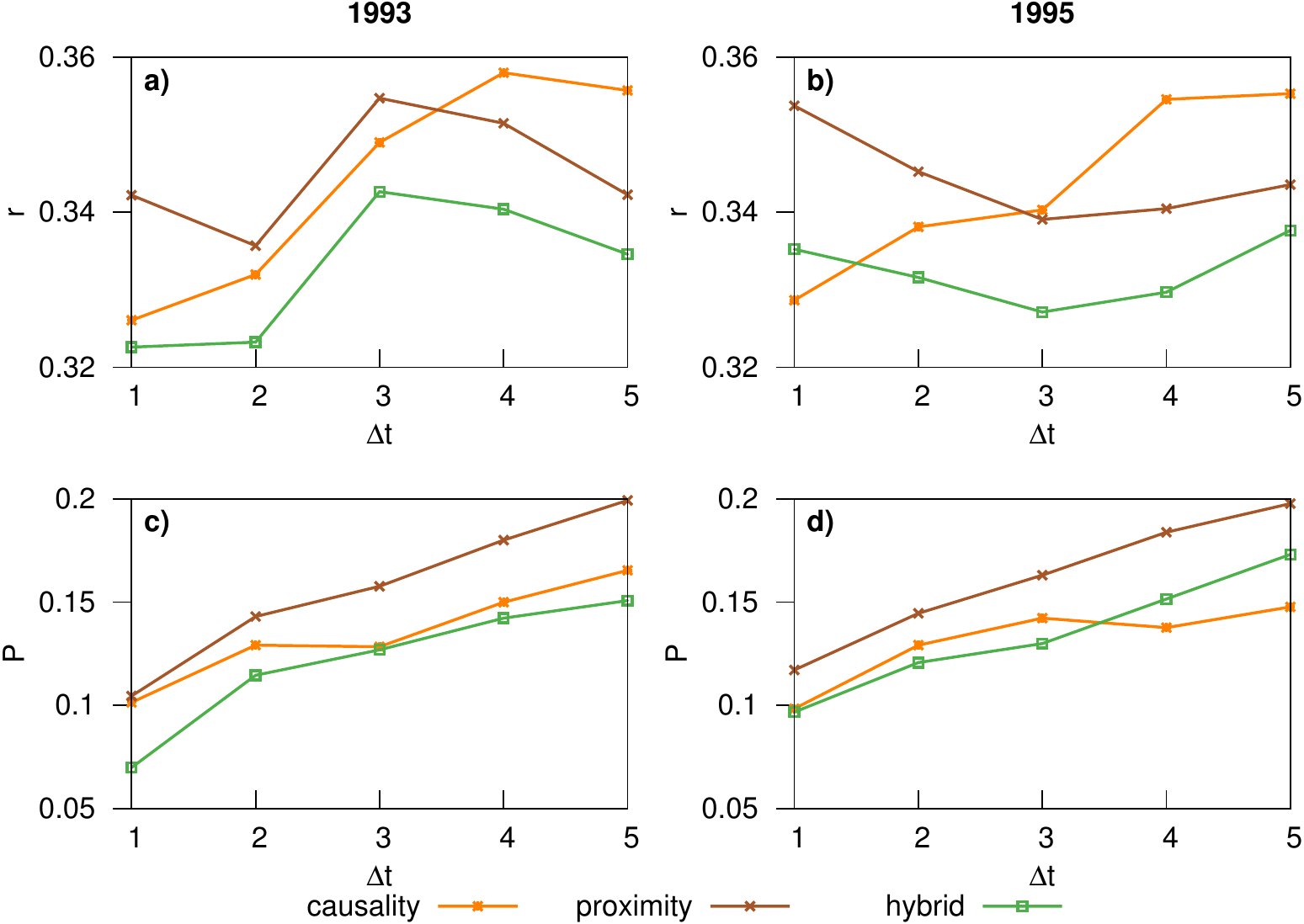}
\caption{We fix here the current year $t$ and do the prediction for year $t+\Delta t$.
a) and b) show the ranking score for causality and proximity relations as well as the hybrid method as a function of $\Delta t$. Panels c) and d)
show the precision as a function of $\Delta t$. For a) and c) the current year is $t=1993$, whereas it is $t=1995$ for b) and d).}
\label{fig:A1}
\end{figure}
In the main text, we predict the new products that countries will export in the next year.
In \prettyref{fig:A1}, we fix a year $t$ and predict the new products that countries will export in year $t+\Delta t$,
but that it is not exporting in year $t$ (in terms of sets, we try to predict $\mathcal{C}^{t+\Delta t} \setminus\mathcal{C}^{t}$).
With respect to the ranking score $r$, causality is optimal when predicting the near future,
but becomes less accurate for higher $\Delta t$. Proximity and hybrid are less sensitive to $\Delta t$. Precision increases with $\Delta t$ for every 
method because the number of new products exported for each country increases with $\Delta t$.

\section{Parameters of causality and proximity relations}
\setcounter{figure}{0}
\setcounter{equation}{0}
\label{app:2}
\begin{figure}[t!]
\centering
\includegraphics[width=1\columnwidth]{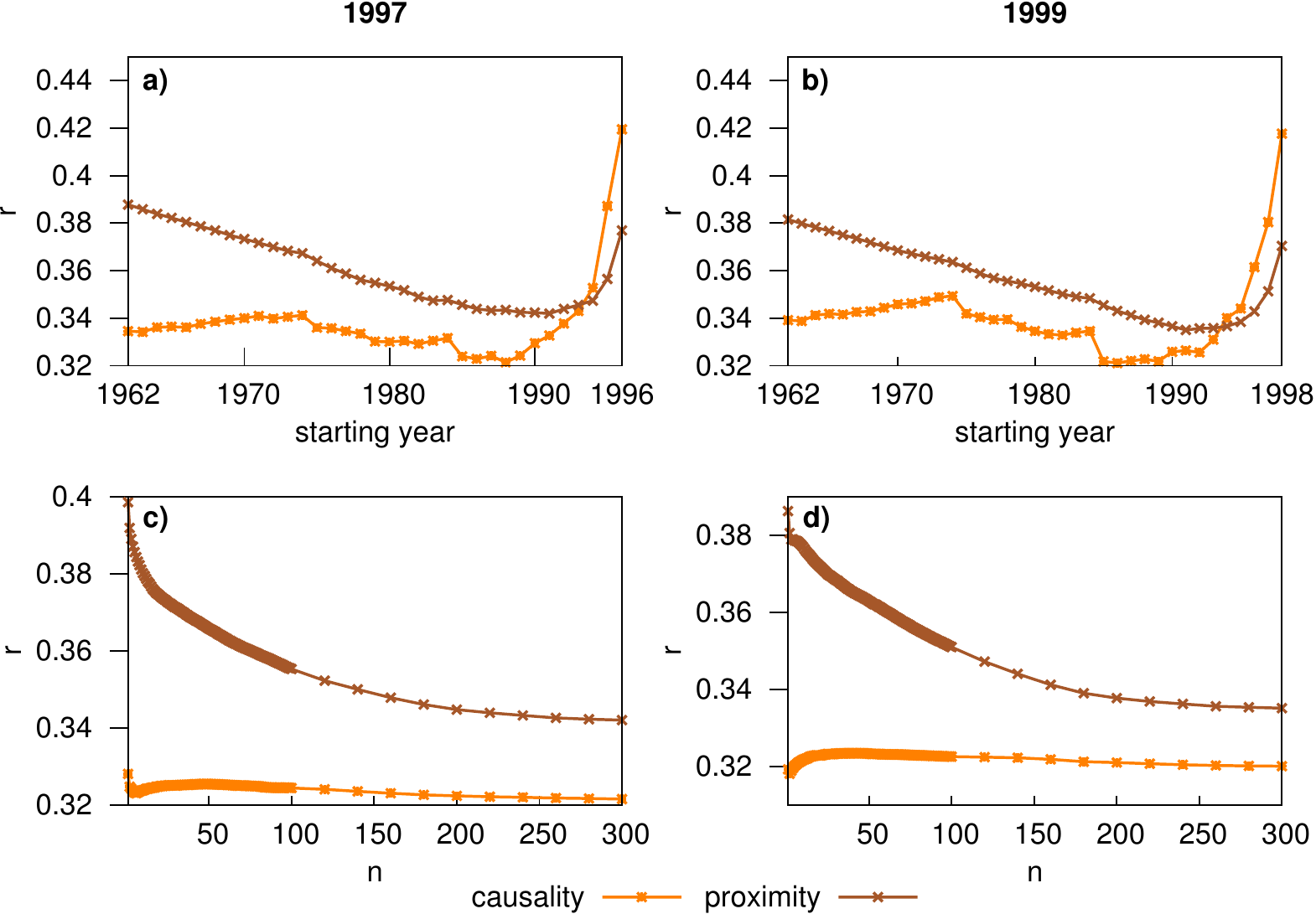}
\caption{a) and b) show the ranking score for causality and proximity relations as a function of the initial year considered to build the relations. c) and d)
show the ranking score as a function of the number of closest products $n$ used in the computation of the closenesses.}
\label{fig:A2}
\end{figure}
Causality and proximity have two parameters each. The first one is the number of years used to compute the resulting values.
The results are shown in 
\prettyref {fig:A2}a and b. For proximity, it is mostly optimal to use data from 1991, which corresponds to the arrival of unified Germany in the dataset.
For causality, while it is generally beneficial to increase the history length, taking data before 1984 results in a worsening of ranking score
due to a change of products' classification original dataset (see Ref. \cite{feenstra2005world}). The causality relations were computed country
by country: for each country we consider that there is a causal relation between $\alpha$ and $\beta$ for country $i$ if $\alpha\in \mathcal{N}_i^t$ and 
$\beta\in \mathcal{C}_i^t$ at least once during the period considered. The causality $\psi_{\alpha\beta}$ is the simply the ratio between the number of causal
relation and the number of countries that export the product at least once.
We can restrain the computation to the $n$ closest products when computing the closeness between a country and a product.
By doing so, we aim to take only the most significant products into account, which is similar to the nearest neighbors typically used
in recommendation algorithms \cite{sarwar2001item}. The results are shown in \prettyref {fig:A2}c and d.
We eventually use the all products into account to compute both proximity and causality. While \prettyref{fig:A2}d
shows that it is not the optimal choice for causality, the difference is rather insignificant (around 0.2\%).

\section{Length of the prediction list}
\label{app:3}
\setcounter{figure}{0}
\setcounter{equation}{0}
\begin{figure}[t!]
\centering
\includegraphics[width=1\columnwidth]{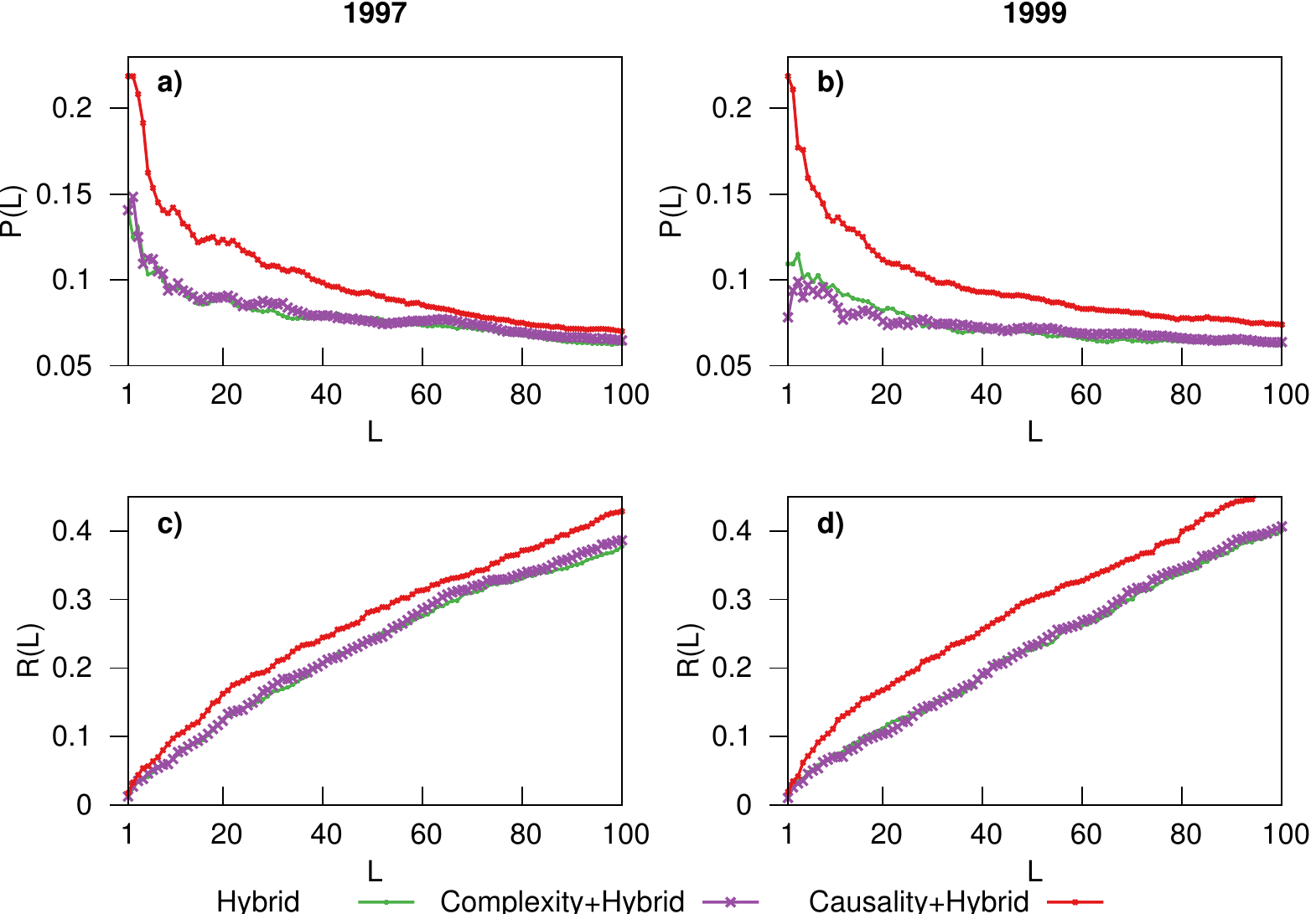}
\caption{Precision $P(L)$ and recall $R(L)$ as a function of the prediction list length $L$ for three different prediction
methods. In panels a) and c) the predictions are done for year 1997, while for panels b) and d) the predictions are done for year 1999.}
\label{fig:A3}
\end{figure}
A prediction method assigns a score to every country-product pair, and generates a prediction list of length $L$ for each country by choosing
the $L$ best-scoring products. The length of the prediction lists is a free parameter, which can be set arbitrarily. We use $L=20$ in the main
test, which reflects the length of a practical prediction list \cite{zhou2010solving}, and it is also close to the mean number of new products 
exported by each country between two consecutive years (approximately 17.2 averaged from 1994 to 2000). For completeness,
we show the dependency of precision and recall on the length of the prediction list in \prettyref{fig:A3}.
Precision decreases with the prediction list length $L$, which shows that the best ranked products
indeed have the highest probability of being exported by a given country in the next year.

\section{The interplay between preferential attachment and Heat diffusion}
\label{app:4}
\setcounter{figure}{0}
\begin{figure}[t!]
\centering
\includegraphics[width=1\columnwidth]{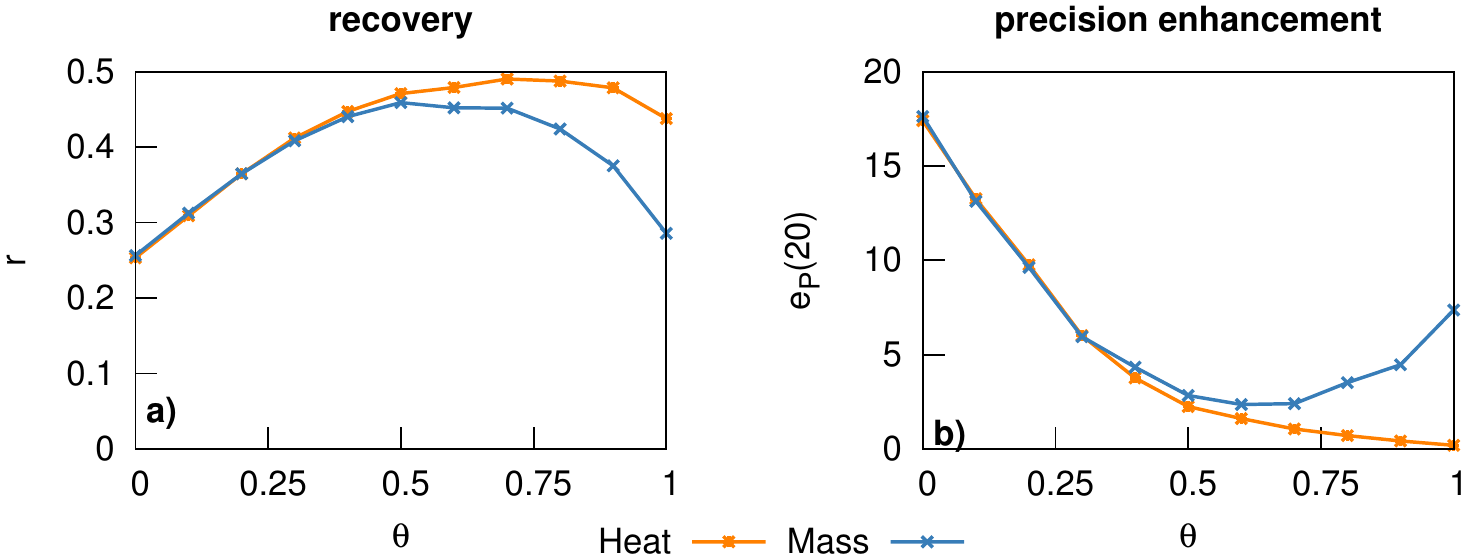}
\caption{Panel a) shows recovery and panel b) enhancement of precision compared to random predictions $e_P(20)$, for both the Mass and Heat 
diffusion recommendation methods as a function of the model parameter $\theta$. The number of users is set to $N=10,000$, number of items to
$M=2,000$ and number of links to $m=100,000$.}
\label{fig:A4}
\end{figure}

\begin{figure}[t!]
\centering
\includegraphics[width=0.5\columnwidth]{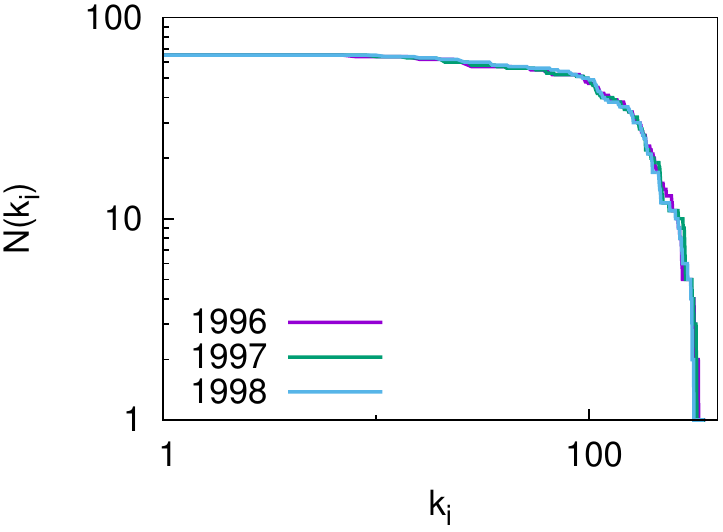}
\caption{The cumulative degree distribution of countries in the country-product network for three different years.}
\label{fig:A5}
\end{figure}
\end{appendices}
We study here a simple model in order to verify that in a network without preferential attachment, it is possible for the Heat diffusion algorithm
to have accuracy comparable with that of Mass diffusion. Our network consists of N users and M items. Each user $i$ has a vector of preferences 
$\mathbf{t_i}$ and each item $\alpha$ has a vector of categories $\mathbf{c_\alpha}$, which correspond to users' tastes. 
An item can either belong to a category or not which corresponds to the elements of category vectors being either one or zero. On the other hand,
a user can either like category, ignore it, or even dislike it, which corresponds to the elements of user preference vectors being $1$, $0$, or $-1$,
respectively. Elements of category vectors are set to 1 or 0 with 50\% probability each. User preference vectors are set to 
zero with 50\% probability, or else to $1$ or $-1$ with equal probability. Links are created one by one between the users and items. 
The user $i$ to which we add a link is chosen among every users with uniform probability, and the item $\alpha$ is chosen with the following rule
\begin{equation}
\tag{D.1}
P(\alpha,i)=\theta \frac{k_\alpha+1}{\sum_\beta k_\beta+1} + (1-\theta) \frac{\mathbf{c_\alpha} \cdot \mathbf{t}_i / O_{\alpha i}}
{\sum_\beta \mathbf{c_\beta} \cdot \mathbf{t_i} / O_{\beta i}}
\end{equation}
where $\theta$ is a parameter to tune the amount of preferential attachment or users’ preferences used in the growth process, and $O_{\alpha i}$
is the number of items $\beta$ which fulfill $\mathbf{c_\beta} \cdot \mathbf{t}_i = \mathbf{c_\alpha} \cdot \mathbf{t}_i$. 
Normalization with $O_{\alpha i}$ enhances the weight of items with high
overlap $\mathbf{c}_{\alpha}\cdot \mathbf{t}_i$ and compensates for the
fact that the number of such items is small. When $\mathbf{c}_{\alpha}\cdot \mathbf{t}_i<0$, the term
proportional to $1-\theta$ is ignored and only the first term contributes to $P(\alpha,i)$.
In total, $m$ links are created in the network. Note that the model setting and parameters are arbitrary and can be modified
to accommodate different behavior of user and items. This elementary model, similar to the agent-based model that was used to
evaluate a news recommendation model~\cite{medo2009adaptive,zhou2011emergence}, is nevertheless sufficient to illustrate the desired
dependency between preferential attachment and recommendation performance.

When the network is built only with preferential attachment ($\theta = 1$), the correlation between the actual degree of
items and their future degree increase is 0.91. When only users’ preferences are used to build the network ($\theta = 0$),
this correlation drops to 0.11. Recommendations obtained with either Heat or Mass diffusion method on artificially created data
are then evaluated using the same approach as the country-product data before. The results are presented in
\prettyref{fig:A4}. In the absence of preferential attachment, Mass and Heat perform similarly in terms of precision and recovery.
By contrast, when preferential attachment is the main driving force ($\theta > 0.75$), Mass clearly outperforms Heat in terms of precision
and recovery. This demonstrates that the weight of preferential attachment in the system's evolution is of crucial importance for the choice 
of the optimal recommendation algorithm. Note that in the model we assume a narrow degree distribution on the user side which is consistent
with the country degree distribution distribution shown in \prettyref{fig:A5} which is also far from the scale-free distributions that are commonly
found in real social and e-commerce data.

\bibliographystyle{model1-num-names}
\bibliography{bibliography}

\end{document}